\documentclass{article}
\usepackage{spconf,amsmath,graphicx}

\usepackage{kotex}
\usepackage{amssymb}
\usepackage{pdfrender}
\usepackage{booktabs}
\usepackage{xcolor}
\usepackage{float}
\usepackage{graphicx}
\usepackage{subcaption}

\usepackage{caption}
\title{Momentum Contrast Speaker Representation Learning}
\name{Jangho Lee$^{1}$ \qquad Jaihyun Koh$^{1,2}$ \qquad Sungroh Yoon$^{1,*}$
\thanks{$^{*}$Correspondence to: Sungroh Yoon $<$sryoon@snu.ac.kr$>$.}}

\address{
$^{1}$Dept. of Electrical and Computer Engineering, Seoul National University, Seoul, Korea \\
$^{2}$Display Electronics Development Team, Samsung Display Corporation, Yongin, Korea \\
{
    \texttt{\{ubuntu,satyricon,sryoon\}@snu.ac.kr}
}
}

\begin{document}
%
\maketitle
\begin{abstract}

Unsupervised representation learning has shown remarkable achievement by reducing the performance gap with supervised feature learning, especially in the image domain.
In this study, to extend the technique of unsupervised learning to the speech domain, we propose the Momentum Contrast for VoxCeleb (MoCoVox) as a form of learning mechanism.
We pre-trained the MoCoVox on the VoxCeleb1 by implementing instance discrimination.
Applying MoCoVox for speaker verification revealed that it outperforms the state-of-the-art metric learning-based approach by a large margin.
We also empirically demonstrate the features of contrastive learning in the speech domain by analyzing the distribution of learned representations.
Furthermore, we explored which pretext task is adequate for speaker verification. 
We expect that learning speaker representation without human supervision helps to address the open-set speaker recognition.
\end{abstract}
\begin{keywords}
speaker recognition, unsupervised learning, self-supervised learning, contrastive learning
\end{keywords}
\section{Introduction}


The dynamic evolution of deep convolutional neural networks has significantly improved the state-of-the-art performance for a wide variety of speech-related tasks such as speaker recognition \cite{nagrani2017voxceleb, chung2018voxceleb2, chung2020defence, huh2020augmentation}, speech enhancement \cite{donahue2018exploring}, and speech separation \cite{ephrat2018looking}. 
Among these tasks, speaker recognition has been extensively studied in recent years. 

\begin{figure}[ht]
    \centering
    \includegraphics[width=\linewidth]{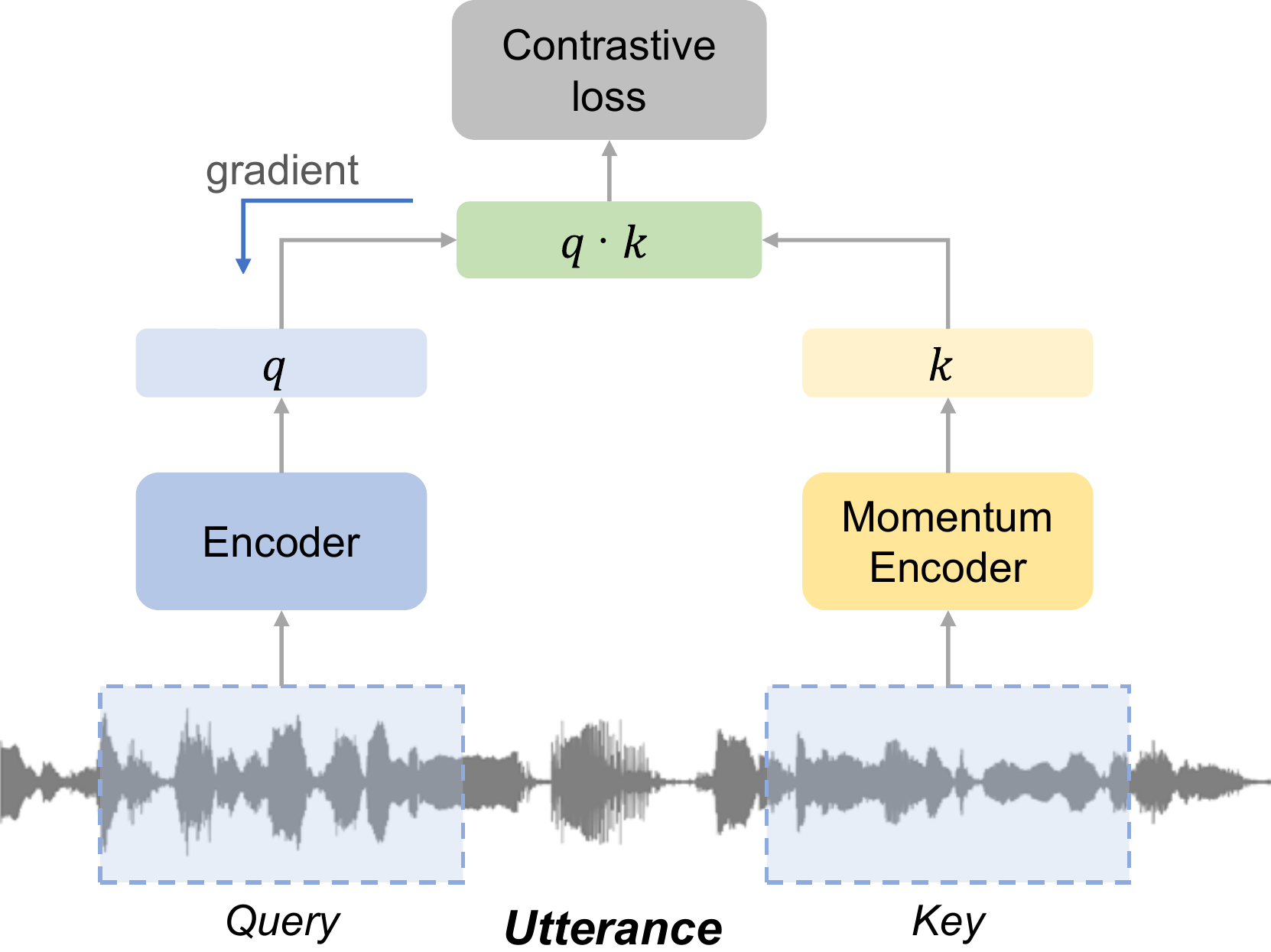}
    \caption{Overview of Momentum Contrast.}
    \label{fig:figure1}
\end{figure}

Speaker recognition can be categorized into two either speaker verification or speaker identification. 
The former determines whether a pair of speech belong to the same identity or not, whereas the latter classifies and assigns an input speech to a specific identity label.
Here, we focus on speaker verification, which is a prominent branch of biometric recognition; thus we have purposed this study to discover a decision boundary that determines whether two groups of speech data were sourced from the same identity or not.
Speaker verification commonly consists of two main components, i.e., a well-trained speaker feature extractor and a data-specific threshold.
In general, a speaker feature extractor is trained with a large-scale speaker recognition dataset such as VoxCeleb \cite{nagrani2017voxceleb, chung2018voxceleb2} and each speaker representation should entail the \textit{identity-specific} information to enhance the speaker verification performance.
The typical way of incorporating identity discriminative information into speaker representation is to modify the objective function since it can be regarded as an open-set problem.
Open-set recognition problem can be solved by metric learning since all test identities are not utilized during training \cite{chung2020defence, liu2017sphereface, wang2018cosface, deng2019arcface, snell2017prototypical}.
The criterion of metric learning is to maximize the \textit{inter-identity} variance and minimize the \textit{intra-identity} variance.
Then, we can find an optimal threshold that yields the best performance on speaker verification benchmark pairs by using techniques such as 10-fold cross-validation.

There has been growing interest in self-supervised learning, which is a form of unsupervised learning that exploits the data itself for supervision.
With this learning scheme, the network parameters are optimized by solving a proxy task formulated with self-supervision \cite{wu2018unsupervised, oord2018representation, chen2020simple, he2020momentum}.
A popular learning paradigm of self-supervised learning is contrastive learning \cite{hadsell2006dimensionality, chen2020simple, he2020momentum} which is used to identify the relationships among data instances by reducing the distance between two samples belonging to the same category (\textit{positive}) while increasing the distance between two samples belonging to different categories (\textit{negative}).
The objective of contrastive learning is in accord with the requirements of the speaker representation for the open-set protocol.

In this study, we propose the MoCoVox for speaker representation learning and attempted to solve a speaker verification by using a self-supervised contrastive learning paradigm.
The reason why a speaker feature extractor should be replaced with such a way of learning is as follows:
First, it is difficult to collect large-scale human-annotated data for fully-supervised speaker recognition tasks.
Secondly, considering the nature of a downstream task, a speaker feature should be distinguishable from features from other identities.
Considering these properties, it is necessary to apply self-supervised learning to the speaker recognition tasks.
Our major contributions can be summarized as follows: 
1) We apply Momentum Contrast on speaker representation learning, and verify the applicability of contrastive learning on the speech domain
2) We demonstrate the effectiveness of Momentum Contrast on the downstream task and outperform the state-of-the-art unsupervised speaker verification performance on the VoxCeleb1 by a large margin.
3) With comprehensive ablation studies, we empirically prove that the properties of contrastive learning are equally applied to speaker representation learning.

\section{Momentum Contrast Learning}

\subsection{Contrastive Learning}
The core principle of contrastive learning \cite{hadsell2006dimensionality} is that the network learns the parameters by pulling neighbors and emphasizing the difference between non-neighbors.
The contrastive learning has evolved with various pretext tasks \cite{zhang2016colorful, bachman2019learning} which is a kind of a proxy task to learn a representation without supervision, effectively.
One of the generally applied pretext tasks is instance discrimination \cite{wu2018unsupervised, bachman2019learning}, which attempts to solve the classification problem by focusing on the image representation itself; specifically, the weight vector of each instance is considered to be a prototype.
Regarding the learning objective, InfoNCE, which was proposed by OOrd et al. \cite{oord2018representation}, it was utilized to contrast the outlier distribution between one positive sample and $k$ negative samples.
The end-to-end update with InfoNCE may be disrupted due to the consistently encoded keys.
Though it is natural to update the parameters in an end-to-end manner, the consistently encoded keys may disrupt the learn \textit{a good representation}.
To solve this problem, Wu et al. \cite{wu2018unsupervised} introduced a memory bank mechanism that contains the pre-calculated representations from the previous iteration. 
As such, in contrastive learning, it is important to contrast the negative representations with respect to the positive representation regardless of class information.

\subsection{Momentum Contrast for Speaker Representation}
We propose a contrastive learning mechanism for speaker representation, called MoCoVox; we also demonstrate its applicability to speech recognition.
As it has a similar architecture to \cite{he2020momentum}, MoCoVox is composed of the following three key components; a query encoder $f_{q}$, momentum encoder $f_{k}$, and dictionary queue, which has a momentum encoder that shares the same architecture with the query encoder.
MoCoVox is designed to \textit{gradually} update a key encoder by maintaining some data samples as a form of representation.
Here, a dictionary inserts the representation of the current mini-batch as the oldest representations are removed following the \textit{queue} operations.
The combination of the three components facilitates InfoNCE optimization as described by Fig.~\ref{fig:figure1}.
When there exists an encoded query $q$ and $K$ encoded keys $k$, InfoNCE can be described by the $(K+1)$-way softmax classification as follows:
\begin{equation}
    \mathcal{L}_{q} = -\log \frac{\mathrm{exp}(q \cdot k^{+} / \tau)} {\sum_{i=0}^{k} \mathrm{exp}(q \cdot k_{i} / \tau)}
    \label{eq:equation1}
\end{equation}
where $q$ represents the output of the query encoder. 
$k^{+}$ indicates the output of the momentum encoder where a key matches to the $q$ while the other values represent the negative keys. 
$\tau$ is a temperature which adjusts the smoothness of the estimated distribution \cite{hinton2015distilling}.
The parameter of momentum encoder is updated by the following equations:
\begin{equation}
    \theta_{k} \leftarrow m\theta_{k} + (1-m)\theta_{q}
        \label{eq:equation2}
\end{equation}
where $m$ is the momentum coefficient. 
Following Eq.~\ref{eq:equation2}, the parameters of momentum encoder $\theta_{k}$ are smoothly tuned rather than those of query encoder.
We can adjust the degree of smoothness by adjusting the momentum coefficient. 
If $m$ approaches $1$, the training process for the query encoder is slow and smooth.
At the same time as the progressive momentum encoder training, the dictionary queue is gradually updated as mentioned earlier.
Unlike the update process for the memory bank approach \cite{wu2018unsupervised}, the parameters of the key encoder are updated according to the momentum-based moving average of the query encoder.

\begin{table}[t!]
    \begin{center}
        \begin{tabular}{l l | c | c | c}

        \toprule
        
        \textbf{Method} & \textbf{Aug.} & \textbf{Arch.} & \textbf{EER (\%)} & \textbf{Vox2} \\
        \midrule
    
        \multicolumn{5} {c} {\bf Fully-supervised baseline} \vspace{3pt} \\
        VoxCeleb1 \cite{nagrani2017voxceleb}    & - & VGG-M         & 7.8 & \\ 
        Voice ID \cite{shon2019voiceid}         & - & TDNN \cite{waibel1989phoneme} & 6.99 & \\
        P \cite{chung2020defence}               & - & ResNet-34     & 4.59 & \checkmark \\
        AP \cite{chung2020defence}              & - & ResNet-34     & 4.29 & \checkmark \\
        
        \midrule
        
        \multicolumn{5} {c} {\bf Self-supervised baseline} \vspace{3pt} \\
        Disent. \cite{nagrani2020disentangled} & - & VGG-M & 22.09 & \checkmark \\ 
        GCL \cite{inoue2020semi} & N $\mid$ R & ResNet-34 & 15.26 & \checkmark \\
        
        \midrule
        
        \multicolumn{5} {c} {\bf No augmentation} \vspace{3pt} \\
        P$^{\dagger}$ \cite{huh2020augmentation}    & - & ResNet-34 & 27.03 & \\
        AP$^{\dagger}$ \cite{huh2020augmentation}   & - & ResNet-34 & 24.18 & \\
        \textbf{MoCoVox}                            & - & ResNet-34 & \textbf{23.60} & \\
        
        \midrule
        
        \multicolumn{5} {c} {\bf Augment one segment} \vspace{3pt} \\ 
        P$^{\dagger}$ \cite{huh2020augmentation}    & N & ResNet-34 & 27.41 & \\
        AP$^{\dagger}$\cite{huh2020augmentation}    & N & ResNet-34 & 23.16 & \\
        \textbf{MoCoVox}                            & N & ResNet-34 & \textbf{18.60} & \\
        
        \midrule
        
        P$^{\dagger}$ \cite{huh2020augmentation}    & N $\mid$ R & ResNet-34 & 26.73 & \\
        AP$^{\dagger}$ \cite{huh2020augmentation}   & N $\mid$ R & ResNet-34 & 23.80 & \\
        \textbf{MoCoVox}                            & N $\mid$ R & ResNet-34 &\textbf{16.11} & \\
        
        \midrule
        
        \multicolumn{5} {c} {\bf Augment both segments} \vspace{3pt} \\
        P$^{\dagger}$ \cite{huh2020augmentation}    & N & ResNet-34 & 20.38 & \\
        AP$^{\dagger}$ \cite{huh2020augmentation}   & N & ResNet-34 & 18.16 & \\
        \textbf{MoCoVox}                            & N & ResNet-34 & \textbf{15.64} & \\
        
        \midrule
        
        P$^{\dagger}$ \cite{huh2020augmentation}    & N $\mid$ R & ResNet-34 & 17.47 & \\
        AP$^{\dagger}$ \cite{huh2020augmentation}   & N $\mid$ R & ResNet-34 & 14.69 & \\
        \textbf{MoCoVox}                            & N $\mid$ R & ResNet-34 & \textbf{13.48} & \\
        
        \bottomrule
        \end{tabular}
    \end{center}
    \caption{Speaker verification performance. Results with $\dagger$ imply the reproduced version with the VoxCeleb1 \cite{nagrani2017voxceleb} dev set. \textbf{P} and \textbf{AP} indicate the Prototypical and Angular Prototypical loss, respectively. 
    \textbf{N} $\mid$ \textbf{R} indicates the augmentation with 75 $\%$ probability with noise and 25 $\%$ probability of room impulse response. The models with \checkmark trained with VoxCeleb2 dev set.}
    \vspace{-1.5em}
    \label{tab:table1}
\end{table}

\vspace{-0.5em}

\section{Experiments}

\subsection{Dataset}
We trained the proposed MoCoVox on the VoxCeleb1 \cite{nagrani2017voxceleb} development set which has 148,642 utterances of 1,211 speakers for network training; 4,874 utterances of 40 speakers are included in the test set that was used to validate the MoCoVox. 
In this study, we only utilized the VoxCeleb1 development set.

It has been proven that data augmentation is an effective way to boost representation quality and improve the performance of downstream tasks in various areas \cite{tian2019contrastive, chen2020simple, he2020momentum}.
To apply augmentation to the data, we utilized MUSAN corpus \cite{snyder2015musan} and room impulse response.

\subsection{Implementation details}
We adopted a Fast ResNet-34 architecture \cite{chung2020defence} and modified the number of output nodes from 512 to 256.
For input to the network, we randomly sampled two segments from the same utterance as a positive pair and sampled two segments from different utterances as a negative pair.
Each speech segment was created by chopping the utterance with 1.8 s, and extracting 40-dimensional log-mel spectrogram with a window length of 25 ms and hop length of 10 ms. 
Following the training process of MoCo \cite{he2020momentum}, we set the dictionary size as 65,536, the temperature $\tau$ as 0.07, and the learning rate as 0.03.

We used the PyTorch \cite{paszke2019pytorch} for our implementation and all experiments were performed on an NVIDIA TESLA V100 GPU.
We will soon make our code and the trained model publicly available.



\subsection{Results}
The primary objective of contrastive learning is to acquire \textit{a good representation} of data by optimizing the pretext tasks to make the downstream tasks easier.
We verified that the pre-trained representation can be transferred to the speaker verification as a downstream task.
In addition, we performed comprehensive experiments to demonstrate that contrastive learning can be effectively applied to speaker representation learning.

\textbf{Evaluation protocol}
To assess the quality of the learned representation, we calculated the equal error rate (EER) that the equilibrium state when the false rejection rate and the false acceptance rate are equal.
We evaluated the EER for the 37,720 utterance pairs, which constitute a random combination of the VoxCeleb1 test dataset.
To determine whether each utterance pair was derived from the same identity, we randomly sampled 10 speech segments from each utterance and extracted the representations for those segments.
Next, calculated 10 $\times$ 10=100 distances using 10 representations from each utterance and took the average for all possible combinations.
Lastly, we identified the optimal threshold, i.e. that which most effectively distinguished the score distribution of positive and negative pairs.

\textbf{Comparison with State-of-the-art} 
To verify the applicability of contrastive learning on the speaker representation learning, we analyzed the speaker verification performance of VoxCeleb1 verification pairs; the results are summarized in Table~\ref{tab:table1}.
We compared the results of our methods to those of various other methods, including an unsupervised learning approach.
For a fair comparison, we retrained the network employed in the current state-of-the-art method \cite{huh2020augmentation}, as it was previously trained on the VoxCeleb2 \cite{chung2018voxceleb2} development set.
A summary of the reproduced performance results is presented in Table~\ref{tab:table1}.
We measured the EER by increasing the degree of augmentation and the number of target segments.
The progressive performance enhancement in EER was observed from 23.60\% to 13.48\% according to adding various noisy sources and increasing the number of segments to which data augmentation is applied.
In particular, when the noise or room impulse response are injected together, the proposed model outperformed the two metric-learning-based objective functions by a large margin.
Thus, we have empirically demonstrated the critical role of data augmentation in contrastive learning \cite{chen2020simple, tian2019contrastive} by reinforcing the discriminative property, providing more negative samples, and attained the competitive results for VoxCeleb1 compared to Huh et al. \cite{huh2020augmentation} that exploits the VoxCeleb2.

\begin{figure}[t]
    \centering
    
    \subfloat[MocoVox]{
        \includegraphics[width=.15\textwidth]{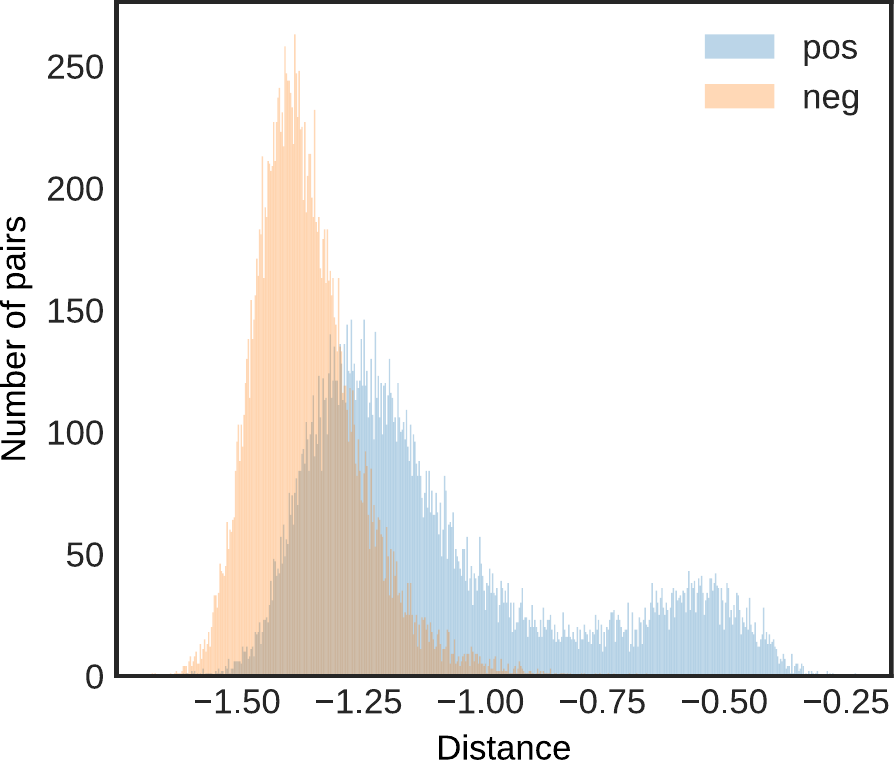}}
    \subfloat[MocoVox (N)]{
        \includegraphics[width=.15\textwidth]{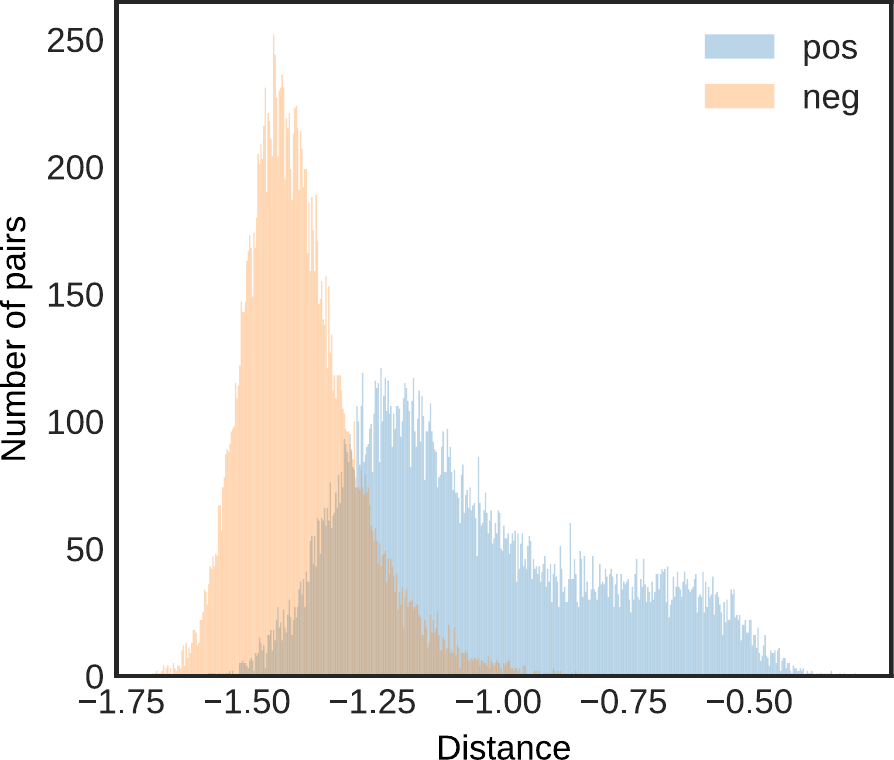}}
    \subfloat[MocoVox (N $\mid$ R)]{
        \includegraphics[width=.15\textwidth]{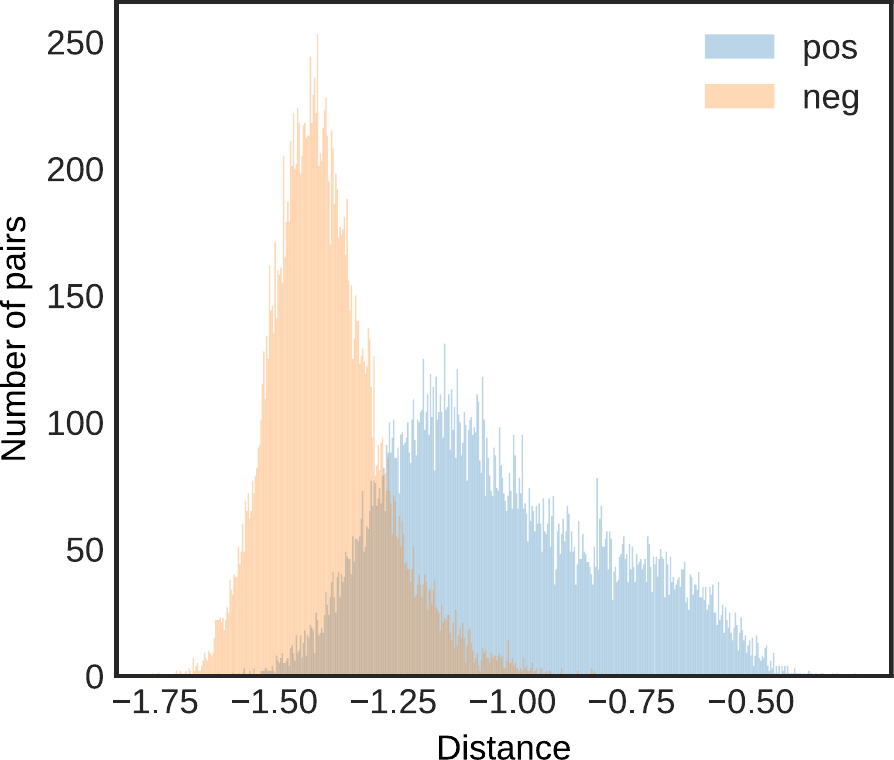}} \\
  
    \vspace{3pt}
    
    \subfloat[P]{
        \includegraphics[width=.15\textwidth]{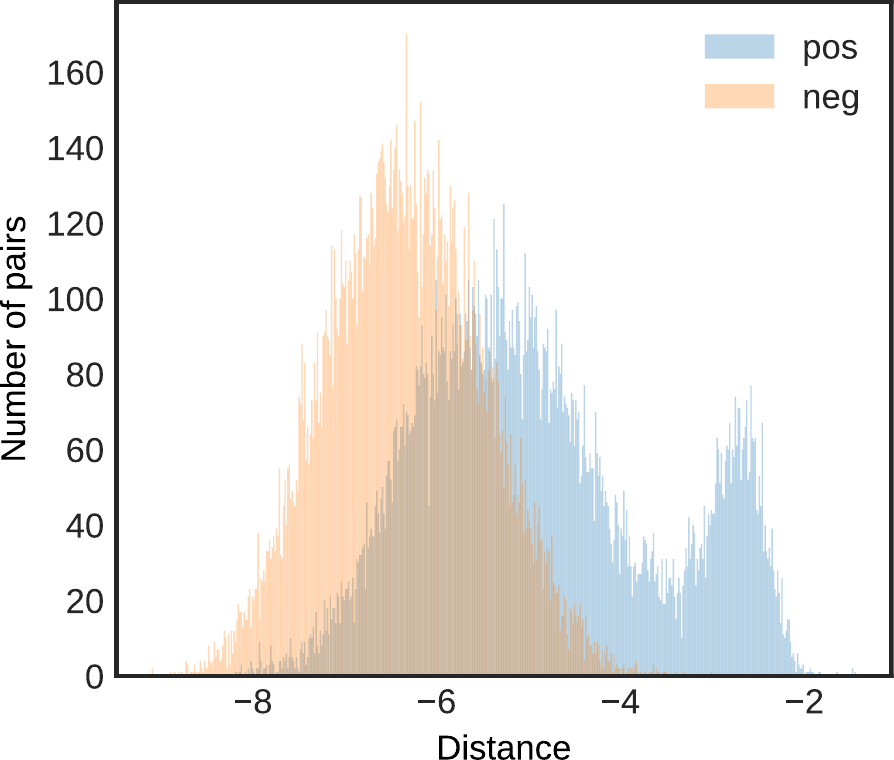}}
    \subfloat[P (N)]{
        \includegraphics[width=.15\textwidth]{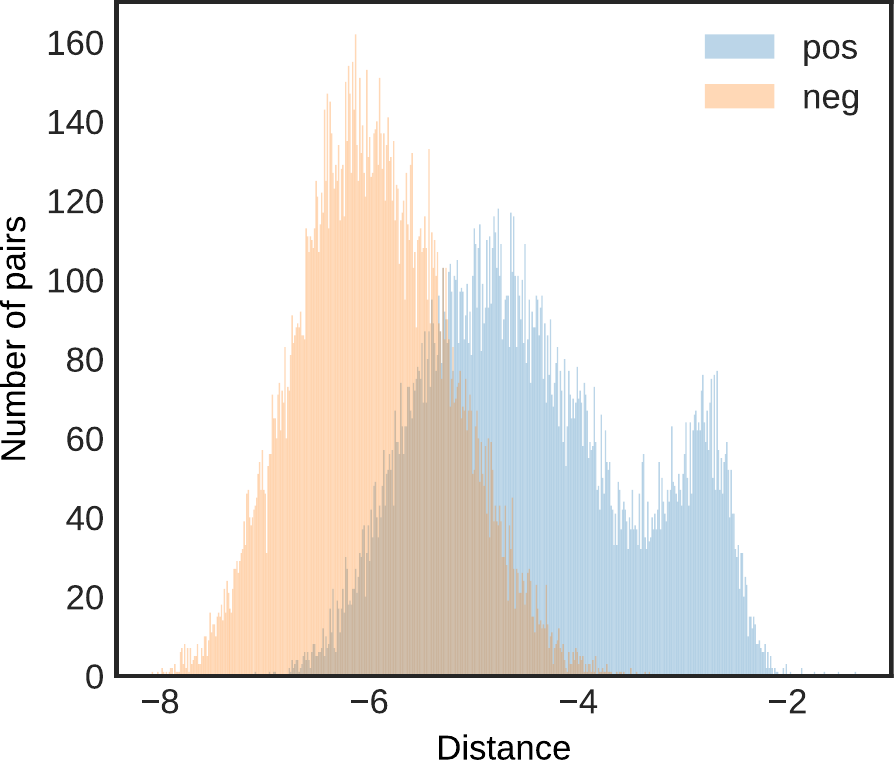}}
    \subfloat[P (N $\mid$ R)]{
        \includegraphics[width=.15\textwidth]{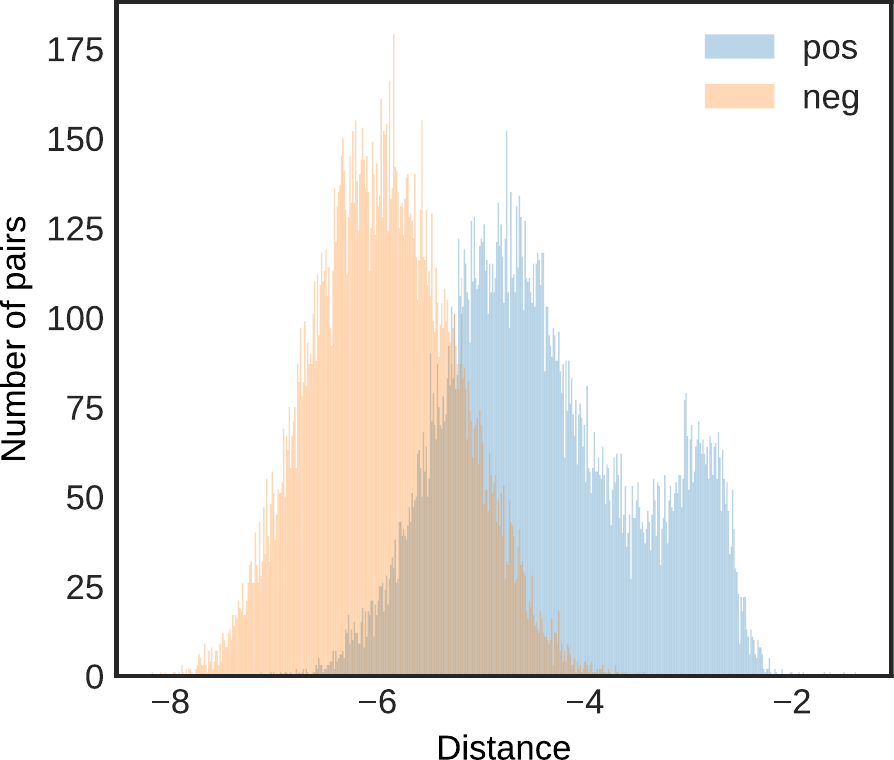}} \\
    
    \vspace{3pt}
    
    \subfloat[AP]{
        \includegraphics[width=.15\textwidth]{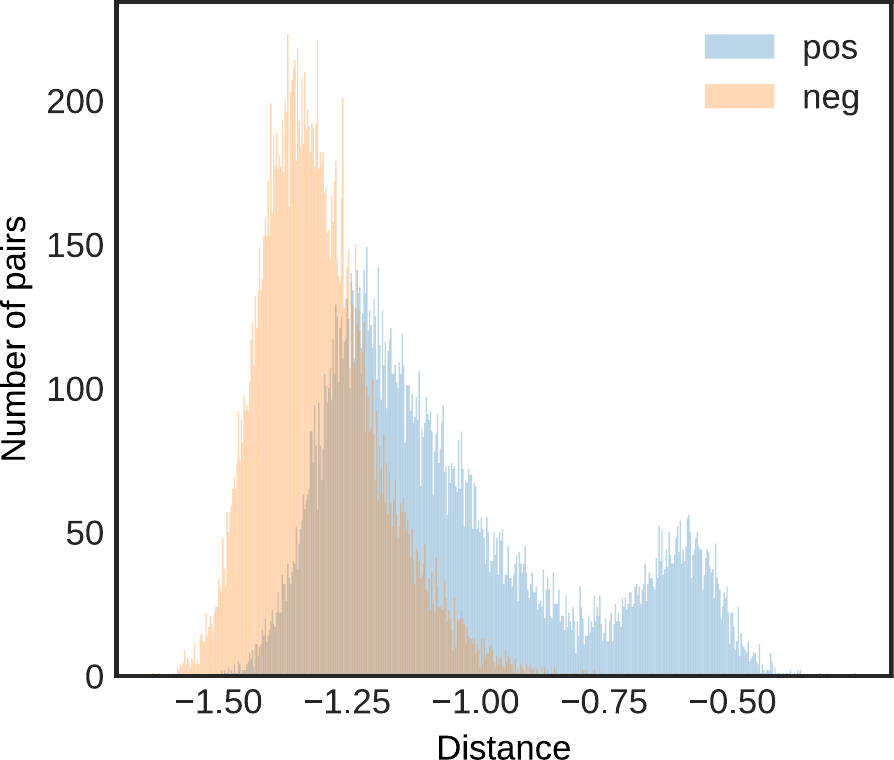}}
    \subfloat[AP (N)]{
        \includegraphics[width=.15\textwidth]{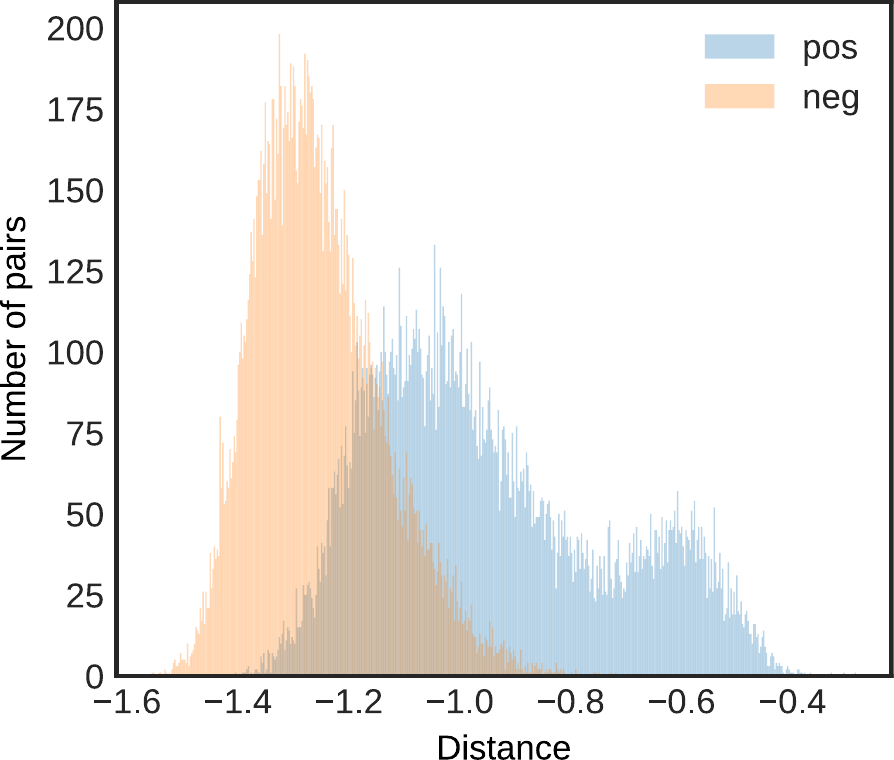}}
    \subfloat[AP (N $\mid$ R)]{
        \includegraphics[width=.15\textwidth]{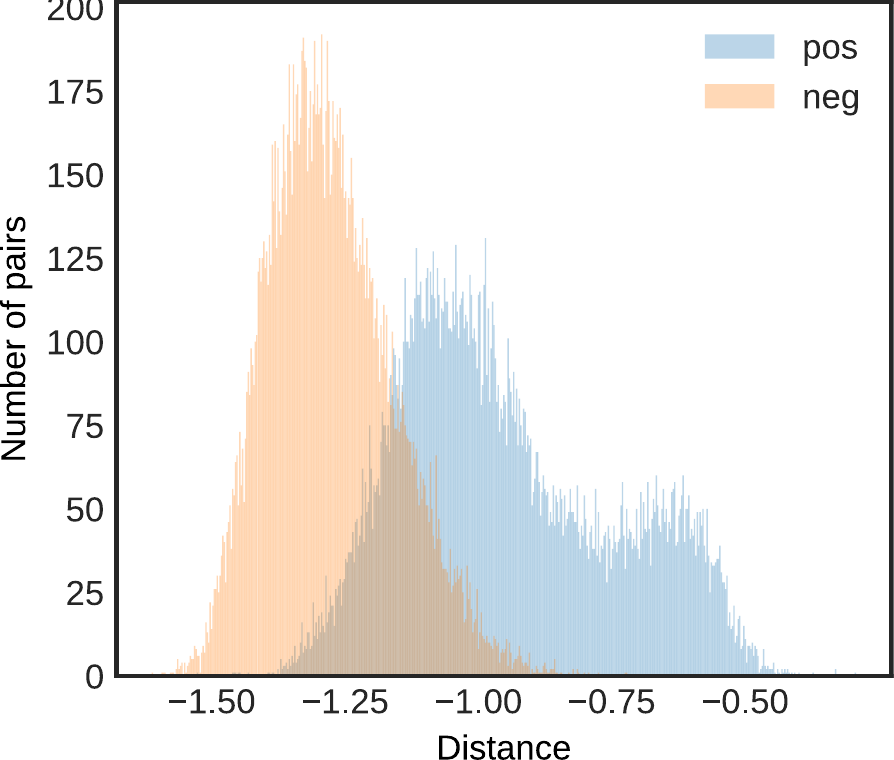}} \\
    
    \caption{Similarity distribution for verification pairs under various augmentations. \textcolor{blue}{Blue} and \textcolor{orange}{orange} histogram represent the similarity between positive pairs and negative pairs, respectively.}
    \label{fig:figure2}
    \vspace{-10pt}
\end{figure}

\textbf{Representation quality} 
We analyzed these empirical results by visualizing the score distribution between positive and negative pairs in Fig.~\ref{fig:figure2}.
For speaker verification, it is important to extract the discriminative representation that can be used to determine whether two speech segments are derived from the same identity.
Based on the shape of the score distribution, the proposed approach has a positive effect on learning representation by reducing the overlapping region (i.e., the false-positive and true-negative samples) for the two distributions.
The distribution shape learned by MoCoVox shows a smooth curve in general.
In particular, in the top-right corner figure, the histogram for the MoCoVox learned with the most strong augmentation accomplishes the best performance with the least overlapped region.
This finding indicates that the proposed learning scheme well learns the data itself by contrasting negative examples.

We experimentally confirmed that the correlation between the size of dictionary look-up and EER.
Under the condition of noisy-only augmentation, we observe that the gradual reduction of EER from 17.99, 17.01, 16.39, 17.06, 16.48, and 15.64 as the size of the dictionary respectively increased from 256, 512, 1024, 2048, 4096, and 65536.
Based on these results, the proposed MoCoVox can be considered to follow the general trend of Momentum Contrast.
Besides, regarding training data quantity, MoCoVox achieved comparable performance with Huh et al. \cite{huh2020augmentation} even though the VoxCeleb1 is approximately seven times smaller than the VoxCeleb2.


\textbf{Pretext tasks} 
We report an experiment in which pretext task yields better results on MoCoVox under the conditions of unsupervised pre-training in Table~\ref{tab:table2}.
We evaluated the individual performances of three pretext tasks, i.e., instance discrimination, and two metric-learning-based tasks.
Instance discrimination, which is the most widely used pretext task is generally employed to learn a representation that classifies individual data instances, without any supervised labels.
Prototypical and angular prototypical loss decrease the distances between representations and the center of the cluster corresponding to the same identity while increase the distances between representations and the center of the cluster corresponding to different identity in a similar way to metric learning.
According to the results presented in Table~\ref{tab:table2}, MoCoVox can contribute to reducing the EER for speaker verification through angular prototypical loss is implemented as a pretext task.

\vspace{-1em}

\section{Conclusion}

\begin{table}[t]
    \begin{center}
        \begin{tabular}{l l | c| c}
        \toprule
        \textbf{Pretext task} & \textbf{Aug.} & \textbf{Arch.} & \textbf{EER (\%)}  \\
        \midrule
    
        \multicolumn{4} {c} {\bf No augmentation} \vspace{3pt} \\
        P \cite{snell2017prototypical}      & - & ResNet-34 & 31.41 \\
        AP \cite{chung2020defence}          & - & ResNet-34 & 28.68 \\
        InsDis \cite{wu2018unsupervised}    & - & ResNet-34 & \textbf{23.60} \\
        
        \midrule
        
        \multicolumn{4} {c} {\bf Augment both segments} \vspace{3pt} \\
        P \cite{snell2017prototypical}      & N & ResNet-34 & 25.20 \\
        AP \cite{chung2020defence}          & N & ResNet-34 & 17.26 \\
        InsDis \cite{wu2018unsupervised}    & N & ResNet-34 & \textbf{15.64} \\
        
        \midrule
        
        P \cite{snell2017prototypical}      & N $\mid$ R & ResNet-34 & 18.71 \\
        AP \cite{chung2020defence}          & N $\mid$ R & ResNet-34 & \textbf{13.04} \\
        InsDis \cite{wu2018unsupervised}    & N $\mid$ R & ResNet-34 & 13.48 \\
        
        \bottomrule
        \end{tabular}
    \end{center}
    \caption{Effect of different pretext tasks.}
    \vspace{-1.5em}
    \label{tab:table2}
\end{table}

In this paper, we propose a speaker representation learning strategy, i.e., MoCoVox, which employs self-supervision to effectively map an input speech segment to a discriminative feature space.
The goal of MoCoVox is to learn \textit{a good representation} for the various downstream tasks similar to other self-supervised learning mechanisms.
We applied the proposed MoCoVox to a speaker verification task; We demonstrates that it outperforms the current state-of-the-art method when trained on the VoxCeleb1.
Furthermore, we observed consistent improvements in verification performance by strengthening the degree of data augmentation.
Through extensive experiments, we empirically demonstrated that the properties of contrastive learning can be effectively applied in the speech domain.

\bibliographystyle{IEEEbib}
\bibliography{manuscript}

\end{document}